\documentclass[10pt,english]{article}

\usepackage[T1]{fontenc}
\usepackage[latin9]{inputenc}
\usepackage[letterpaper]{geometry}
\usepackage{color}
\usepackage{float}
\usepackage{graphicx}
\usepackage{setspace}
\usepackage{esint}
\usepackage[numbers]{natbib}
\makeatletter
\newcommand{\lyxaddress}[1]{
\par {\raggedright #1
\vspace{1.4em}
\noindent\par}
}

\@ifundefined{date}{}{\date{}}
\usepackage{babel}

\makeatother

\usepackage{babel}
\begin{document}

\title{\textbf{The Generic Unfolding of a Biomimetic Polymer}\\
 \textbf{during Force Spectroscopy}\\
 \vspace{4ex}
 }

\author{\textit{Aviel Chaimovich}\textsuperscript{\textit{1a}}\textit{,
Christian Leitold}\textsuperscript{2,3}\textit{, and Christoph Dellago}\textsuperscript{\textit{3b}}\\
 \vspace{-4ex}
 }
\maketitle

\lyxaddress{}

\lyxaddress{\begin{center}
\textsuperscript{\textit{1}}\textit{Department of Chemical and Biological
Engineering, Drexel University, Philadelphia, PA 19104}\\
 \textsuperscript{\textit{2}}\textit{Department of Chemical and Biomolecular
Engineering, University of Illinois at Urbana-Champaign, IL 61801}\\
 \textsuperscript{\textit{3}}\textit{Faculty of Physics, University
of Vienna, 1090 Wien, Austria}\\
 \vspace{2ex}
 \textsuperscript{\textit{a}}\textit{aviel.chaimovich@drexel.edu}\\
 \textsuperscript{\textit{b}}\textit{christoph.dellago@univie.ac.at}\\
 \vspace{2ex}
 
\par\end{center}}
\begin{abstract}
With the help of force spectroscopy, several analytical theories aim
at estimating the rate coefficient of folding for various proteins.
Nevertheless, a chief bottleneck lies in the fact that there is still
no perfect consensus on how does a force generally perturb the crystal-coil
transition. Consequently, the goal of our work is in clarifying the
generic behavior of most proteins in force spectroscopy; in other
words, what general signature does an arbitrary protein exhibit for
its rate coefficient as a function of the applied force? By employing
a biomimetic polymer in molecular simulations, we focus on evaluating
its respective activation energy for unfolding, while pulling on various
pairs of its monomers. Above all, we find that in the vicinity of
the force-free scenario, this activation energy possesses a negative
slope and a negative curvature as a function of the applied force.
Our work is in line with the most recent theories for unfolding, which
suggest that such a signature is expected for most proteins, and thus,
we further reiterate that many of the classical formulae, that estimate
the rate coefficient of the crystal-coil transition, are inadequate.
Besides, we also present here an analytical expression which experimentalists
can use for approximating the activation energy for unfolding; importantly,
it is based on measurements for the mean and variance of the distance
between the beads which are being pulled. In summary, our work presents
an interesting view for protein folding in force spectroscopy. 
\end{abstract}

\lyxaddress{}

\lyxaddress{\begin{center}
\vspace{4ex}
 
\par\end{center}}

\pagebreak{} 

\section{\textit{Introduction \label{Sec:Introduction}}}

\begin{onehalfspace}
The folding of proteins, from expanded coils to collapsed crystals,
has been a main research focus for several decades by now \citep{DillWeikl_ARB2008,ThirumalaiHyeon_ARB2010}.
One particularly successful route for studying such a process has
been via force spectroscopy: \textcolor{black}{By pulling on a pair
of residues of a certain protein, various properties associated with
the chain (e.g., the distance between the two beads) can be measured
as a function of the applied force} \citep{GreenleafBlock_ARBS2007,Bustamante_ARB2008}.
Importantly, the peculiarly rapid folding of a typical protein, to
a very low-entropy native state from a very high-entropy chaotic state,
is still an unresolved conundrum, and thus, many studies of force
spectroscopy especially focus on the kinetic properties of this transition,
in both its forward and backward directions \citep{SchwaigerRief_Nature2002,CecconiMarqusee_Science2005,GM0Fernandez_PNAS2009,RP0Fernandez_CP2016}. 

\begin{figure}[H]
\begin{centering}
\includegraphics[scale=0.5]{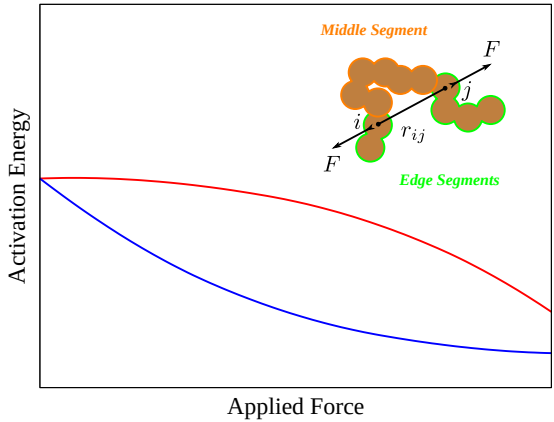} 
\par\end{centering}
\caption{\textit{\label{Fig:Sketch} }A sketch of the force spectroscopy for
a generic unfolding of a biomimetic polymer. The graph plots the two
main trends suggested for the activation energy in terms of the applied
force (on an arbitrary pair of beads): While both curves have a negative
slope throughout their domain, they differ in their curvature, with
the blue being convex and the red being concave. Besides, we also
have here a schematic which expresses the main notation for the force
spectroscopy in our work. The force $F$ is applied between monomers
$\left\{ i,j\right\} $ in the pulling direction of their pairwise
distance $r_{ij}$. In our nomenclature, we call the portion of the
chain that is between $i$ and $j$ the middle segment, and the two
portions to the left and right of the chain are named the edge segments.
In this specific sketch, $N=12$, with $i=2$ and $j=4$, meaning
that it is an asymmetric situation in force spectroscopy; the corresponding
$\Delta n$ (i.e., the relative separation in index space) is $7$. }
\end{figure}

By collecting data at finite forces, while also invoking various analytical
theories, a chief aim of the scientific community is in estimating
the rate coefficient of the (intrinsic) folding process in the force-free
scenario \citep{Makarov_JCP2016}. In an Arrhenius-like manner, the
analytical theories generally assume that the activation energy for
unfolding is essentially equal, within a constant, to the logarithm
of the corresponding rate coefficient. On a phenomenological level,
Bell popularized in the biological community a formula, which presumes
that the Newtonian work of pulling is entirely absorbed by the activation
energy \citep{Bell_Science1978}. As an improvement, Dudko et al.\ presented
another analytical expression that notably accounts for the stochastic
nature of the folding process via Kramers approach \citep{DudkoSzabo_PRL2006};
we sketch the respective activation energy versus the relevant applied
force as the blue curve in Fig.\ \ref{Fig:Sketch}. Nevertheless,
such a kinetic theory is still deficient in describing the empirical
behavior of all proteins. Specifically in the vicinity of the force-free
scenario, the molecular simulations of Best et al.\ for ubiquitin
showed that the activation energy, as a function of the pulling force,
exhibits a negative slope and a negative curvature \citep{BestDudko_JPCB2008},
while the experimental study of Jagannathan et al.\ for another protein
domain reaffirmed this trend \citep{JagannathanMarqusee_PNAS2012};
we sketch it as the red curve in Fig.\ \ref{Fig:Sketch}. Note that
although the Dudko formula also has a negative slope throughout its
domain, it possesses the opposite sign for the curvature. It has been
argued that the main source for this discrepancy stems from the fact
that this analytical theory assumes that the distance between the
two beads, on which the force is applied, is always the reaction coordinate
for the crystal-coil transition \citep{Makarov_JCP2016}. By assuming
that there is just one other component to the reaction coordinate
besides the bead-bead distance (e.g., the number of contacts a protein
makes with itself), it has been demonstrated that various options
for the slope and curvature signs can be obtained \citep{SuzukiDudko_PRL2010,ZhuravlevThirumalai_PNAS2016}.
Besides, a generalized approach for the force-free scenario naturally
lies in a Taylor expansion for the activation energy: The linear term
is basically equivalent with the Bell formula, while the quadratic
term accounts for the ``compliance'' (i.e., an effective compressibility,
which is formally defined via the force derivative of the bead-bead
distance) \citep{HuangBoulatov_PAC2010}. 
\end{onehalfspace}

At the most basic level, one may wonder what signs, for the slope
and for the curvature, do most proteins exhibit in their force functionality.
In other words, given an arbitrary protein, do we know which trend
of Fig.\ \ref{Fig:Sketch} it is likely to possess? An ideal model
for answering this question is the biomimetic polymer of Taylor et
al., which, most notably, is composed of identical beads \citep{TaylorBinder_PRE2009,TaylorBinder_JCP2009}.
By employing the Wang\textendash Landau algorithm \citep{WangLandau_PRL2001}
for molecular simulations, it was shown that this homogeneous generic
polymer mimics the prominent signature of heterogeneous biological
polymers: With a particular choice of parameters, this biomimetic
chain exhibits a discontinuous (finite-size) phase transition between
a collapsed crystal and an expanded coil with no presence of a globule
state \citep{TaylorBinder_PRE2009,TaylorBinder_JCP2009}. Importantly,
an ensuing study focused on the force-free kinetics of a slight variant
of this polymer, examining many order parameters (e.g., the gyration
radius) as possibilities for the reaction coordinate of the crystal-coil
transition \citep{LeitoldDellago_JCP2014,LeitoldDellago_JPCM2015}.
Above all, the total potential energy of the chain, in comparison
with all other order parameters examined, obtained the best representation
for the reaction coordinate of the folding process. Importantly, the
absolute value of this order parameter roughly corresponds with the
number of (random) contacts the generic chain makes with itself. This
is rather analogous with many kinetic studies which assume that the
number of (native) contacts a biological chain makes with itself is
a good reaction coordinate for protein folding \citep{SunOnuchic_BJ2014,BerkovichBerne_JPCB2017}.
\vspace{2ex}
 
\begin{onehalfspace}
Our current work aims at complementing the understanding of the generic
crystal-coil transition during force spectroscopy. At the most basic
level, via the perturbation theory of Zwanzig \citep{Zwanzig_JCP1954},
we apply a force on an arbitrary monomeric pair of the biomimetic
polymer studied by the molecular simulations of Refs.\ \citep{LeitoldDellago_JCP2014,LeitoldDellago_JPCM2015}.
Reminiscent of the various theories for force spectroscopy \citep{Makarov_JCP2016},
we aim at clarifying transition kinetics by focusing here on free
energies: The main assumption of our study is that the total potential
energy of the polymer is an adequate approximation for the reaction
coordinate of the crystal-coil transition. We consequently show, for
generic unfolding, that its activation energy in the force-free limit
consistently exhibits a negative slope and a negative curvature as
a function of the applied force, which is in line with findings for
ubiquitin \citep{BestDudko_JPCB2008}, as well as for other protein
domains \citep{JagannathanMarqusee_PNAS2012}. Our work thus suggests
that most proteins exhibit such a a functionality in force spectroscopy,
and that other trends are exceptions to this general rule.

\vspace{2ex}

\end{onehalfspace}

\section{\textit{Polymer System \label{Sec:System} }}

\begin{figure}[H]
\begin{centering}
\includegraphics[scale=0.5]{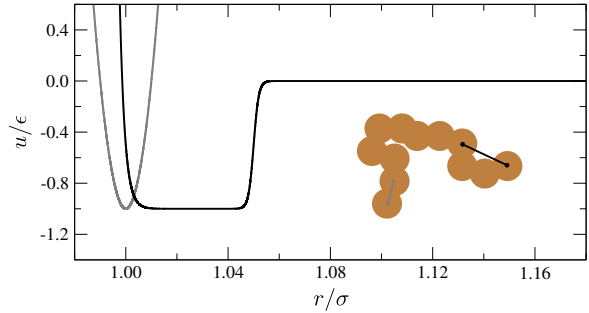} 
\par\end{centering}
\caption{\textit{\label{Fig:Model} }The model of the biomimetic polymer of
identical beads. Besides the schematic of the chain, we also present
here the pairwise potential in terms of the pairwise distance. The
black curve is relevant for non-neighboring beads: It is the singularity-free
version of a wall-like core with a square-like well given by Eq.\ \ref{Eq:Nonbond}.
The gray curve is the harmonic bond of Eq.\ \ref{Eq:Bond}; for clarity,
it is shifted downwards by unity. }
\end{figure}

\begin{onehalfspace}
As mentioned earlier, the basis for our biomimetic polymer is the
model of Taylor et al.\ \citep{TaylorBinder_PRE2009}; a schematic
of an arbitrary configuration of such a chain is given in the inset
of Fig.\ \ref{Fig:Model}. We specifically use the singularity-free
version of this model devised in Ref.\ \citep{LeitoldDellago_JCP2014}
for purposes of studying the kinetic behavior of the polymer (its
equilibrium phenomena is essentially identical with that of the original
model). Depicted in Fig.\ \ref{Fig:Model} as the black curve, the
pairwise potential $u$ between non-neighboring beads is a function
of their pairwise distance $r$, and it is given by \citep{LeitoldDellago_JCP2014}:

\begin{equation}
u\left(r\right)=\frac{1}{2}\epsilon\left[\exp\left[-\alpha\left(r/\sigma-1\right)\right]+\tanh\left[\alpha\left(r/\sigma-\lambda\right)\right]-1\right]\label{Eq:Nonbond}
\end{equation}
The spatial parameter $\sigma$ is the width of the wall-like core,
and the energetic parameter $\epsilon$ is the depth of the square-like
well; our entire study is made dimensionless in terms of these two.
The dimensionless $\lambda$ controls the diameter of the latter:
We fix it at $\lambda=1.05$, since this is a value which can attain
the aforementioned crystal-coil transition (i.e., a chief characteristic
of protein folding); note that this value makes the attractive well
very narrow compared to the repulsive core. The tuning parameter $\alpha$
regulates the smoothness of this function, and we fix it at $\alpha=500$.
Furthermore, depicted in Fig.\ \ref{Fig:Model} as the gray curve,
the harmonic bonds of this polymer are governed by \citep{LeitoldDellago_JCP2014}:

\begin{equation}
u\left(r\right)=\frac{1}{2}\epsilon\kappa\left(r/\sigma-1\right)^{2}\label{Eq:Bond}
\end{equation}
The dimensionless spring constant $\kappa$ is set to $20,000$. Besides,
we set the number of monomers $N$ to $128$. Note that $\alpha\rightarrow\infty$,
together with $\kappa\rightarrow\infty$, retrieves the original model
of Taylor et al.\ \citep{TaylorBinder_PRE2009}. \vspace{2ex}
 
Given the positions of all monomers $r^{N}$, the total potential
energy of the chain $U$ is defined by the pairwise summation of Eq.\ \ref{Eq:Nonbond},
together with that of Eq.\ \ref{Eq:Bond}. Once a force is applied
between an arbitrary pair of beads $i$ and $j$, the configurational
Hamiltonian of the system $H_{F}$ is given by:

\begin{equation}
H_{F}\left(r^{N}\right)=U\left(r^{N}\right)-Fr_{ij}\label{Eq:Hamiltonian}
\end{equation}
Here, $r_{ij}$ is the scalar distance between the two relevant monomers
(its trivial dependence on $r^{N}$ is omitted for clarity), while
$F$ is the scalar force, defined in the pulling direction of $r_{ij}$.
Note that for the force-perturbed system, $H_{F}$ is the total potential
energy, not just $U$. Our corresponding notation for the bead indices
is the following: The index $i$ or $j$ is the ordinal number of
the monomer counting from the leftmost or rightmost edge, respectively.
As a clarification, the symmetric scenario corresponds with $i=j$;
for example,$\left\{ i,j\right\} =\left\{ 1,1\right\} $ means the
force is applied on the edge beads, while $\left\{ i,j\right\} =\left\{ \frac{1}{3}\left(N+1\right),\frac{1}{3}\left(N+1\right)\right\} $
is the case in which the polymer is partitioned into three equal segments.
We frequently call the polymeric portion between the $\left\{ i,j\right\} $
monomers as the middle segment, while the other two portions are referred
to as the edge segments. We also define their corresponding separation
in index space:

\begin{equation}
\Delta n=N-\left(i+j\right)+1\label{Eq:IndexSeparation}
\end{equation}
This notation is sketched in Fig.\ \ref{Fig:Sketch}.

Importantly, note that specific values of the configurational functions
which appear in Eq.\ \ref{Eq:Hamiltonian} can be used as order parameters
(e.g., a bead-bead distance $r_{ij}$). In our work, the most important
order parameter which we examine is $U$. Above all in the force-free
study of Taylor et al., it was shown that the total potential energy
can clearly discriminate between the collapsed and expanded states
of their biomimetic polymer \citep{TaylorBinder_PRE2009}. As we alluded
to earlier, $-U/\epsilon$ is fairly representative of the number
of contacts that the monomers make with each other, since Eq.\ \ref{Eq:Nonbond}
is almost binary in its functionality, with its dimensionless version
being essentially $1$ or $0$ depending on whether a contact is established
or removed, respectively. In turn, the collapsed state has many contacts
with a large $\left|U\right|$, and the expanded state has few contacts
with a small $\left|U\right|$ (a representative schematic can be
found in Fig.\ \ref{Fig:ProbabilityDistributionU}). In consideration
of our dimensionless study, we frequently refer to $\left|U\right|$
(sometimes even colloquially to $U$ itself) as the ``contacts-number''.
On a relevant note, we generally use $\mathcal{A}$ and $\mathcal{B}$
as labels for the crystal and coil states, respectively, while a double
dagger $\ddagger$ is employed for referring to the transition region
between these two metastable phases.

\vspace{2ex}

\end{onehalfspace}

\section{\textit{Molecular Simulations of the Force-Free Scenario \label{Sec:WangLandau} }}

\begin{onehalfspace}
Examining both the crystal and coil phases of the polymer, especially
the folding process, is rather cumbersome: A regular molecular simulation
would spend most of its time in one of the phases, and it would rarely
make the transition between the two. This bottleneck can be overcome
via the Wang\textendash Landau algorithm, which iteratively evaluates
the density of states of the system, while also generating a respective
flat histogram for the ensemble of configurations in terms of the
total Hamiltonian energy as the order parameter \citep{WangLandau_PRL2001}.
Still, the Wang\textendash Landau strategy is quite expensive computationally,
and thus in our specific study, instead of performing the algorithm
for various force values, as well as for various bead pairs, we essentially
execute it only for the special case of $F=0$; for treating force
spectroscopy, we employ the perturbation theory of Zwanzig \citep{Zwanzig_JCP1954},
which is thoroughly discussed in a later section below. Regardless,
the molecular simulations of the biomimetic polymer in the force-free
scenario are essentially the basis for all of the data presented here.

\begin{figure}[H]
\begin{centering}
\includegraphics[scale=0.5]{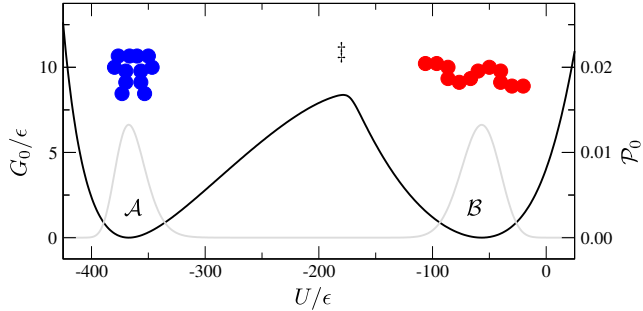} 
\par\end{centering}
\caption{\textit{\label{Fig:ProbabilityDistributionU} }The force-free transition
between the crystal and coil phases at the temperature of coexistence
(i.e., $kT=0.438\epsilon$). Plotted in terms of the ``contacts-number''
as the order parameter, the black curve is the free energy (on the
left ordinate), while the gray curve is the probability distribution
(on the right ordinate); the arbitrary constant of the former is chosen
so that the metastable phases have a value of zero. The left and right
schematics correspondingly depict the collapsed and expanded states.}
\end{figure}

In the absence of a force, the total Hamiltonian energy is identical
with the total potential energy (i.e., $H_{0}=U$), and in turn, the
Wang\textendash Landau approach, via the density of states $\Omega_{0}$,
conveniently provides the corresponding (canonical) probability distribution
$\mathcal{P}_{0}$,

\begin{equation}
\mathcal{P}_{0}\left(U\right)=\Omega_{0}\left(U\right)\exp\left[-U/kT\right]/Z_{0}\label{Eq:ProbabilityDistributionU0}
\end{equation}
in which $Z_{0}$ is the (canonical) partition function; $k$ is the
Boltzmann constant, while $T$ is the temperature of interest. Within
an arbitrary constant, the respective free energy $G_{0}$ is given
by the logarithm of the above;

\begin{equation}
G_{0}\left(U\right)=-kT\ln\Omega_{0}\left(U\right)+U+\mathrm{cnst.}\label{Eq:FreeEnergyU0}
\end{equation}
Throughout this publication, we choose the arbitrary constant so that
the folded state has zero energy. We also restrict our current work
to the coexistence temperature between the crystal and coil phases
at $F=0$: This corresponds with $kT=0.438\epsilon$ \citep{LeitoldDellago_JCP2014}.
We show the corresponding free energy (as the black curve), together
with the probability distribution (as the gray curve), in Fig.\ \ref{Fig:ProbabilityDistributionU}.
It is now obvious that the two metastable phases are separated by
a rare transition region. Throughout our work, we define the crystal
$U^{\mathcal{A}}$ and the coil $U^{\mathcal{B}}$ as the values of
the order parameter which correspond with the minima in the free energy.
Analogously, we also denote $U^{\ddagger}$ as the value of the ``contacts-number''
associated with the maximum in the free energy. For this force-free
scenario, $U^{\mathcal{A}}\approx-367\epsilon$ and $U^{\mathcal{B}}\approx-57\epsilon$,
while at the same time, $U^{\ddagger}\approx-179\epsilon$. \vspace{2ex}
 
The probability distribution of Eq.\ \ref{Eq:ProbabilityDistributionU0},
together with all the configurations generated by the molecular simulations,
can be employed for reweighting any property of the polymer at $U$
into the equivalent property at $T$. The reweighting of an average
of an arbitrary configurational function $\psi\left(r^{N}\right)$
can be numerically determined via the following integration over the
order parameter:

\begin{equation}
\left\langle \psi\right\rangle _{0}=\int\left\langle \psi\right\rangle _{0}^{U}\mathcal{P}_{0}\left(U\right)dU\label{Eq:Reweight0}
\end{equation}
The brackets here mean ensemble averaging: The presence of the index
$U$ corresponds with averaging at that particular value of the ``contacts-number'';
if this index is absent, canonical averaging at $T$ is implied. Note
that there is another version of this expression based on the delta
function: $\left\langle \delta\left(U\right)\psi\right\rangle _{0}=\left\langle \psi\right\rangle _{0}^{U}\mathcal{P}_{0}\left(U\right)$;
$\delta\left(U\right)$, in the brackets, is an abbreviated notation
for the familiar delta function, $\delta\left(\hat{U}\right)=\delta\left(\hat{U}-U\left(r^{N}\right)\right)$,
with its dependence on the positions of all beads implied.

We briefly describe our current implementation of the Wang\textendash Landau
algorithm, which is equivalent with the one of Ref.\ \citep{LeitoldDellago_JCP2014}.
The molecular simulations evolve via Monte Carlo moves: Besides the
standard single-particle displacement, there are four special types
of polymeric moves, which are randomly executed half of the time (on
average). One pass of the simulations consists of $2N=256$ moves.
Using the already converged density of states calculated in Ref.\ \citep{LeitoldDellago_JCP2014},
we examine the order parameter $U$ over two windows, $\left[-430\epsilon,-70\epsilon\right]$
and $\left[-70\epsilon,130\epsilon\right]$, using a bin size of $1\epsilon$.
We record the positions of all monomers every $10^{5}$ passes, which
ensures that any two configurations are completely decorrelated from
each other. Conventional rules for modification factors, as well as
for histogram flatness, are employed. We perform $16$ sets of this
algorithm in parallel, and we eventually combine all of their data
together. For any given $U$, we harvest about $4,000$ uncorrelated
configurations of the polymer.

\vspace{2ex}

\end{onehalfspace}

\section{\textit{Pairwise Distance between Two Monomers \label{Sec:Distance}}}

\begin{onehalfspace}
The probability distribution for $U$ of Fig.\ \ref{Fig:ProbabilityDistributionU},
which is directly generated by the Wang\textendash Landau algorithm,
can be reweighted to a probability distribution for any other order
parameter. In particular, let us consider an arbitrary pairwise distance
$r_{ij}$ (at this point, there is still no force between beads $i$
and $j$). Remember that this is an order parameter which is frequently
assumed as a reaction coordinate in many kinetic theories for protein
folding \citep{DudkoSzabo_PRL2006}. We consequently substitute the
delta function $\delta\left(r_{ij}\right)$ for $\psi$ in Eq.\ \ref{Eq:Reweight0},
attaining the following:

\begin{equation}
\mathcal{P}_{0}\left(r_{ij}\right)=\int\mathcal{P}_{0}^{U}\left(r_{ij}\right)\mathcal{P}_{0}\left(U\right)dU\label{Eq:ProbabilityDistributionR0}
\end{equation}
Note that $\mathcal{P}$ with an extra index $U$ (i.e., $\mathcal{P}_{0}^{U}$)
corresponds to a conditional distribution at that value of the ``contacts-number'';
the absence of such an index means that a probability distribution
is canonical at $T$. Employing the many Wang\textendash Landau configurations,
we numerically evaluate this probability distribution: For a given
bead pair $\left\{ i,j\right\} $, we construct a histogram in terms
of their $r_{ij}$ at a specific $U$, and by running across this
entire order parameter, while also using its (canonical) probability
distribution $\mathcal{P}_{0}\left(U\right)$, we correspondingly
obtain the (canonical) probability distribution for the pairwise distance
$\mathcal{P}_{0}\left(r_{ij}\right)$. \vspace{2ex}
 
\begin{figure}[H]
\begin{centering}
\includegraphics[scale=0.5]{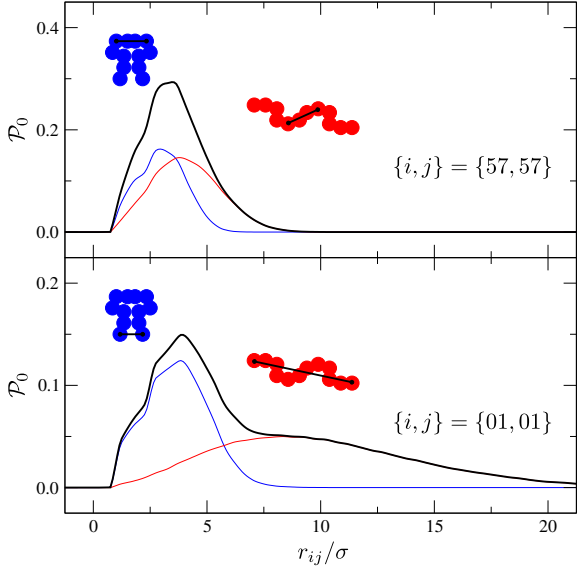} 
\par\end{centering}
\caption{\textit{\label{Fig:ProbabilityDistributionR} }The probability distribution
for the pairwise distance between a bead pair in the force-free scenario.
The panels only differ in terms of their specific monomeric pair,
with the top being for $\left\{ i,j\right\} =\left\{ 57,57\right\} $
and the bottom being for $\left\{ i,j\right\} =\left\{ 01,01\right\} $;
notice that the ordinate-axis differs by a factor of two. The black
curve is the probability distribution across all of $U$. The colored
curves are probability distributions for restricted regions of the
``contacts-number'': Specifically represented by their corresponding
schematics, the blue and red curves are for the crystal and coil states,
respectively. }
\end{figure}

In Fig.\ \ref{Fig:ProbabilityDistributionR}, we present this probability
distribution as the black curve for two sets of bead pairs, with the
top panel for $\left\{ i,j\right\} =\left\{ 57,57\right\} $ and the
bottom panel for $\left\{ i,j\right\} =\left\{ 01,01\right\} $; realize
that the top range is twice as much as the bottom range. The noticeable
difference between the two panels is associated with the overall monotonic
relationship between the bead separation in index space and the bead
separation in real space: Larger $\Delta n$ generally involves larger
$r_{ij}$, and thus, the probability distribution has higher values
in the top panel than in the bottom panel. Besides this numerical
discrepancy, the common aspect in both panels is that this $\mathcal{P}$
does not at all exhibit two clearly separated hills (compare with
the probability distribution of Fig.\ \ref{Fig:ProbabilityDistributionU}).
This means that $r_{ij}$ is not an adequate order parameter for distinguishing
between the crystal and coil phases, which in turn implies that it
cannot be the reaction coordinate!

Let us clarify the situation here. We can partition the overall $\mathcal{P}$
into its two main contributions, one around the crystal basin, $U/\epsilon=\left[-402,-302\right]$
and one around the coil basin $U/\epsilon=\left[-114,-14\right]$;
these boundaries for $U$ are chosen so that each metastable phase
corresponds with a $G_{0}$ that is roughly within $5kT$ of its basin.
Labeled by the colored schematics, the separate crystal and coil contributions
to the entire probability distribution are given respectively as the
blue and red curves in Fig.\ \ref{Fig:ProbabilityDistributionR}.
Due to the low probability of finding the system outside of the stable
basins, the sum of the two curves is almost identical with the black
curve. It is now clear that there is a significant overlap between
the separate probability distributions of the crystal and coil phases:
While $r_{ij}$ is quite small for the crystal and quite large for
the coil, this plain two-body property can fluctuate considerably
for each of the metastable phases, and it cannot describe the collective
multi-body transition of the entire polymer. This overlap is consequently
the source for the inadequacy of $r_{ij}$ as an order parameter. 
\end{onehalfspace}

In line with our generic polymer, other recent studies also suggest
that the pairwise distance between arbitrary beads cannot be generally
presumed as the reaction coordinate for the crystal-coil transition
\citep{Makarov_JCP2016}. Nevertheless, this may not be necessarily
the case for protein folding. Since a biological chain folds in a
very unique manner, its probability distribution for a pairwise distance
is much sharper, and it can also involve many other basins. In turn,
a given protein may have a certain $r_{ij}$ that cleanly distinguishes
between the collapsed and expanded phases, while also playing a key
role in the reaction mechanism. 

\begin{onehalfspace}
\vspace{2ex}

\end{onehalfspace}

\section{\textit{Application of a Force between Two Monomers \label{Sec:Perturbation}}}

\begin{onehalfspace}
We again emphasize that the focal part of the current work is in mimicking
force spectroscopy, specifically examining the influence of an applied
force $F$ (between an arbitrary bead pair $\left\{ i,j\right\} $)
on the free energy $G_{F}$ of our generic chain, in terms of the
``contacts-number'' $U$ as the order parameter. While for lattice
systems, such a calculation may be feasible by the direct application
of the Wang-Landau algorithm \citep{LS0Binder_JCP2014}, this may
not be the case for our polymeric model. In turn, rather than performing
the Wang\textendash Landau algorithm for many values of $F$ (with
numerous sets of $\left\{ i,j\right\} $), we go about this task via
a rather computationally inexpensive route, which is based on the
perturbation theory of Zwanzig \citep{Zwanzig_JCP1954}. With the
force-free chain being the reference system in the Zwanzig formalism,
the applied force is the perturbation parameter which changes the
total Hamiltonian energy from $H_{0}$ to $H_{F}$. Given an arbitrary
configurational function $\psi\left(r^{N}\right)$, Zwanzig perturbation
means the following for Eq.\ \ref{Eq:Hamiltonian}:

\begin{equation}
\left\langle \psi\right\rangle _{F}=\left\langle \psi\exp\left[Fr_{ij}/kT\right]\right\rangle _{0}/\left\langle \exp\left[Fr_{ij}/kT\right]\right\rangle _{0}\label{Eq:ZwanzigF}
\end{equation}
In line with the notation of the zero index above (e.g., $\left\langle \psi\right\rangle _{0}$),
the $F$ index denotes the value of the force which corresponds with
the appropriate ensemble average. Substituting the delta function
$\delta\left(U\right)$ for $\psi$ in the above yields:

\begin{equation}
\mathcal{P}_{F}\left(U\right)=\mathcal{P}_{0}\left(U\right)\left\langle \exp\left[Fr_{ij}/kT\right]\right\rangle _{0}^{U}/\left\langle \exp\left[Fr_{ij}/kT\right]\right\rangle _{0}\label{Eq:ProbabilityDistributionUF}
\end{equation}
Note that we also invoke here the infinitesimal version of Eq.\ \ref{Eq:Reweight0}
with $\psi=\exp\left[Fr_{ij}/kT\right]$. We consequently have the
respective free energy,

\begin{equation}
G_{F}\left(U\right)=G_{0}\left(U\right)-kT\ln\left\langle \exp\left[Fr_{ij}/kT\right]\right\rangle _{0}^{U}+\mathrm{cnst.}\label{Eq:FreeEnergyUF}
\end{equation}
with an arbitrary constant (independent of $U$ but not of $F$),
which is again set at a value that ensures $G_{F}=0$ for the folded
state. We emphasize here that the entire formalism (e.g., Eqs.\ \ref{Eq:ProbabilityDistributionU0},
\ref{Eq:Reweight0}, \ref{Eq:ZwanzigF}, \ref{Eq:FreeEnergyUF}, etc.)
is completely universal. As such, all expressions above can be employed
in examining any protein; the only unique feature that one must consider
is the specific model used for $H_{0}$, and this may also influence
the exact choice for the order parameter of the crystal-coil transition. 

\vspace{2ex}
 
\begin{figure}[H]
\begin{centering}
\includegraphics[scale=0.5]{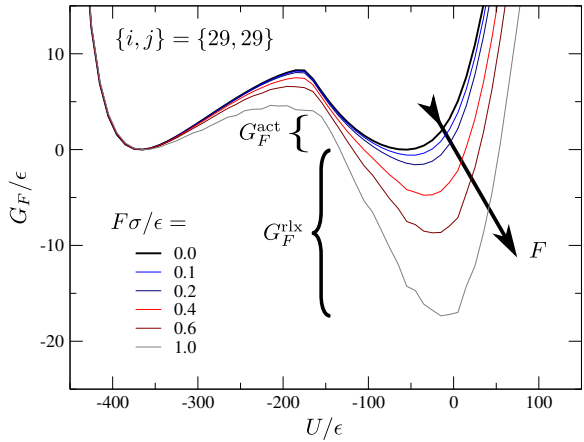} 
\par\end{centering}
\caption{\textit{\label{Fig:FreeEnergyUF} }The free energy as a function of
the ``contacts-number'', while varying the magnitude of the applied
force. Each curve corresponds to a different value of the force. The
black one is actually for the force-free scenario, and it is identical
with its counterpart in Fig.\ \ref{Fig:ProbabilityDistributionU}.
The other colored curves correspond with the polymer under force spectroscopy:
The arrows mean that the value of the force increases, and in this
graph, it is always applied between monomers $\left\{ 29,29\right\} $.
Besides, we schematically depict here the activation and relaxation
energies associated with the gray curve (i.e., the case of $F=1.0\epsilon/\sigma$). }
\end{figure}

We now specifically apply the above formalism on our polymer. We numerically
loop over all $U$ for a particular bead pair $\left\{ i,j\right\} $:
At a specific $U$, we essentially run over all of the relevant (force-free)
Wang\textendash Landau configurations, averaging for them $\exp\left[Fr_{ij}/kT\right]$,
simultaneously for many values of $F$. Note that because the exponential
factor is an increasing monotonic function of $Fr_{ij}$, this perturbation
term naturally generates noticeable statistical uncertainty for the
unfolded state around $F\approx1\epsilon/\sigma$ (in comparison with
the folded state around $F\approx0\epsilon/\sigma$); for alleviating
this issue, we average, upon the completion of the protocol, all of
our data by lumping $U$ in bins of $10$. Besides, in line with other
sections of this publication, we only perform this procedure at the
coexistence temperature of the force-free scenario; realize, of course,
that upon (finite) force spectroscopy, the crystal and coil metastable
phases are not at coexistence anymore. Such a Zwanzig protocol has
negligible computational cost compared to the Wang\textendash Landau
algorithm, and thus it is rather trivial to perform the Zwanzig calculation
for many different bead pairs. 

Consider the specific monomeric pair $\left\{ 29,29\right\} $: This
is a symmetric scenario for force spectroscopy, in which the two edge
segments are (roughly) a quarter of the total polymeric length, while
the middle segment is the remaining (roughly) half of the polymer.
For this set of beads, we present its free energy as a function of
the ``contacts-number'' in Fig.\ \ref{Fig:FreeEnergyUF}, while
varying the magnitude of the applied force. The black line is for
the force-free scenario, which is basically identical with the analogous
curve of Fig.\ \ref{Fig:ProbabilityDistributionU}; all other colors
correspond with different values of a (finite) force. As expected,
increasing the magnitude of the pulling force (represented by the
arrow) makes the transition from the crystal to the coil more and
more favorable: The change in the free energy between the two metastable
states increases (in terms of its absolute value), and this occurs
concurrently with a decreasing barrier for unfolding. \vspace{2ex}
 
\begin{figure}[H]
\begin{centering}
\includegraphics[scale=0.5]{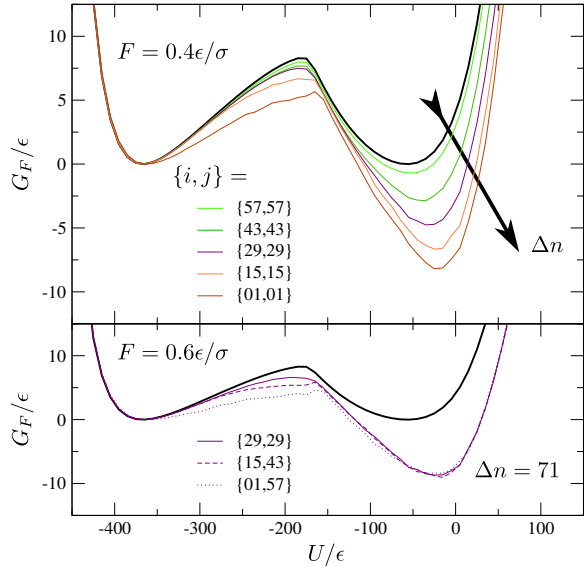} 
\par\end{centering}
\caption{\textit{\label{Fig:FreeEnergyUij} }The free energy as a function
of the ``contacts-number'', with the force applied on various monomeric
pairs. The top panel examines $F=0.4\epsilon/\sigma$, and it is only
for symmetric scenarios, in which the length of the middle portion,
in index space, is varied, while its midpoint is fixed at the center
of the polymer; the arrow corresponds with an increase in this length,
and the color of a curve determines the respective value. The bottom
panel examines $F=0.6\epsilon/\sigma$, and it is only for asymmetric
scenarios, in which the middle segment of the polymer is always the
same length in index space (i.e, $71$), while the corresponding location
of its midpoint is varied. }
\end{figure}

Let us now fix the force at a constant value, while varying the beads
on which it is applied. In the top panel of Fig.\ \ref{Fig:FreeEnergyUij},
with $F=0.4\epsilon/\sigma$, we specifically examine symmetric scenarios
of force spectroscopy: The two edge segments are always equal in length,
while we just vary the relative separation, in index space, between
the two monomers (i.e., $\Delta n$). Interestingly, varying this
parameter gives a trend analogous with the one mentioned above in
the context of Fig.\ \ref{Fig:FreeEnergyUF}: With increasing $\Delta n$,
as well as with increasing $F$, $G_{F}$ shifts to lower values across
its entire order parameter (considering our choice for the arbitrary
constant). This correspondence of $\Delta n$ with $F$ stems in the
monotonic relationship between the bead separation in index space
and the bead separation in real space, discussed in the context of
Fig.\ \ref{Fig:ProbabilityDistributionR}: Since $r_{ij}$ has the
same functionality as $F$ in the Zwanzig factor (i.e., $\exp\left[Fr_{ij}/kT\right]$),
increasing this pairwise distance should have the same general effect
as amplifying the magnitude of the applied force. On the other hand,
the bottom panel of Fig.\ \ref{Fig:FreeEnergyUij} presents several
asymmetric scenarios for $F=0.6\epsilon/\sigma$; while the two edge
segments are not of the same length here, they are of the same family
in the sense that $\Delta n$ is fixed between them. Interestingly,
for such a family of curves, the difference in free energy between
the two metastable phases is essentially unperturbed, while the barrier
for unfolding noticeably changes. Overall, it is clear that there
are subtle effects on $G_{F}$ across its order parameter, depending
on which specific beads the force is applied.

\vspace{2ex}

\end{onehalfspace}

\section{\textit{Moments of the Pairwise Distance \label{Sec:DistanceMoments}}}

\begin{figure}[H]
\begin{centering}
\includegraphics[scale=0.5]{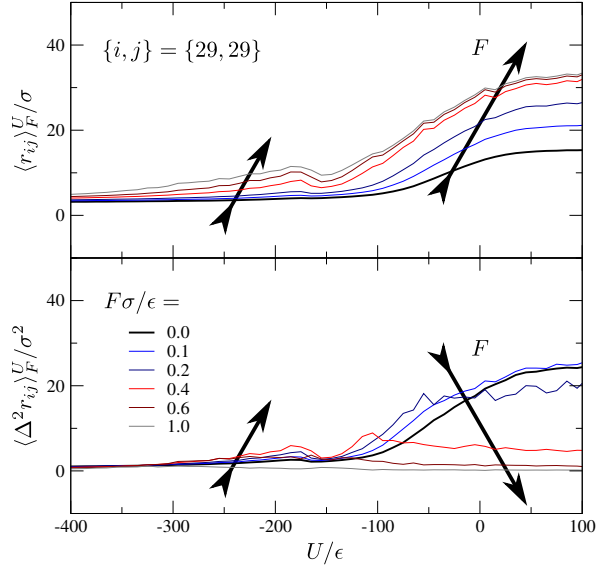} 
\par\end{centering}
\caption{\textit{\label{Fig:DistnaceMomentUF} }The moments of the pairwise
distance as a function of the ``contacts-number''. The top panel
is for the mean, and the bottom panel is for the variance. The color
coding here for the applied force, on beads $\left\{ 29,29\right\} $,
is identical with the one in Fig.\ \ref{Fig:FreeEnergyUF}, and again,
the arrows denote the general trend for an increasing force.}
\end{figure}

\begin{onehalfspace}
Let us now switch back our focus to the other order parameter that
we explore in this publication (the reason for this will become clear
in the next section). Specifically, we examine moments of the pairwise
distance between an arbitrary monomeric pair; this essentially corresponds
with the moments of the probability distributions of Fig.\ \ref{Fig:ProbabilityDistributionR},
except that we now vary the applied force. We focus on the first and
second moments, which allow us to obtain mean and variance of the
pairwise distance. The moments can be calculated by setting $\psi$
equal to different powers of $r_{ij}$ in Eq.\ \ref{Eq:ZwanzigF},
together with Eq.\ \ref{Eq:Reweight0}. We implement this computation
in the numerical script described in the previous section. We plot
in the top and bottom panels of Fig.\ \ref{Fig:DistnaceMomentUF}
the respective mean $\left\langle r_{ij}\right\rangle _{F}^{U}$ and
variance $\left\langle \Delta r_{ij}^{2}\right\rangle _{F}^{U}$ as
a function of the ``contacts-number''; we introduce here $\Delta r_{ij}=r_{ij}-\left\langle r_{ij}\right\rangle _{F}^{U}$,
with the dependence of the delta symbol on the indices (e.g., $F$)
being implicit. The coding here is analogous with the one in Fig.\ \ref{Fig:FreeEnergyUF}:
The different colors are for the same values of the force, which is
again applied on $\left\{ i,j\right\} =\left\{ 29,29\right\} $. 
\end{onehalfspace}

For the $F=0$ case (i.e., the black curves), the trend is the same
for both moments: As expected, they are both low for the (rigid) crystal,
and they are both high for the (floppy) coil. While this function
is almost monotonic throughout, there is a slight inflection in the
transition region (i.e., at $U\approx-200\epsilon$), which corresponds
with the location of the barrier in the free energy. Taylor et al.\ noticed
an analogous behavior for the gyration radius \citep{TaylorBinder_JCP2009},
and overall, this may correspond to a subtlety in the nucleation process:
As the biomimetic polymer transitions from a crystal to a coil, there
is an instance during which various of its contacts are broken, yet
its size contracts. On a related note, the string analysis of Ref.\ \citep{LeitoldDellago_JPCM2015}
observed a switch in the main contribution to the reaction coordinate
(i.e., from ``contacts-number'' to ``crystallinity'') around the
same location of $U$. \textbackslash{}

Once a force is applied, we observe completely different behavior
between the mean and the variance. The relationship for $\left\langle r_{ij}\right\rangle _{F}^{U}$
is quite simple, with a stronger force yielding a larger mean. Nevertheless,
the situation is rather complex for $\left\langle \Delta r_{ij}^{2}\right\rangle _{F}^{U}$,
being notably non-monotonic: At a high ``contacts-number'' (e.g.,
$-U/\epsilon\approx300$), the variance generally increases with $F$,
yet at a low ``contacts-number'' (e.g., $-U/\epsilon\approx-100$),
the variance generally decreases with $F$. Besides, it seems that
both moments reach an asymptote at $F=1.0\epsilon/\sigma$, since
they do not change much beyond this value. Here is how one can understand
the situation. As we pull stronger and stronger on the crystal state,
it disentangles, with the beads getting farther apart, while fluctuating
more on its periphery. On the other hand, as we pull stronger and
stronger on the coil state, the beads very easily stretch out between
each other, and being very stretched, they do not have much of a possibility
for fluctuations.\vspace{2ex}
 
\begin{onehalfspace}
\vspace{2ex}

\end{onehalfspace}

\section{\textit{Approximations for Force Spectroscopy \label{Sec:Taylor}}}

\begin{onehalfspace}
For various reasons, one may be interested in estimating the approximate
effect of force spectroscopy on a certain polymer. For example, one
can solely perform a molecular simulation of a force-free scenario
of a protein, while subsequently predicting the free energy for a
finite force; conversely, experiments can be done at a finite force,
while estimating the free energy for the force-free case.

As such, we correspondingly formulate a Taylor series for $G_{F}\left(U\right)$
in the vicinity of $F=0$. In consideration of Eq.\ \ref{Eq:FreeEnergyUF},
here is its first-order derivative,

\[
\frac{\partial}{\partial F}G_{F}\left(U\right)=-\frac{\left\langle r_{ij}\exp\left[Fr_{ij}/kT\right]\right\rangle _{0}^{U}}{\left\langle \exp\left[Fr_{ij}/kT\right]\right\rangle _{0}^{U}}
\]

\begin{equation}
\left[\frac{\partial}{\partial F}G_{F}\left(U\right)\right]_{F=0}=-\left\langle r_{ij}\right\rangle _{0}^{U}\label{Eq:FreeEnergyUF1}
\end{equation}
and here is its second-order derivative,

\[
\frac{\partial^{2}}{\partial F^{2}}G_{F}\left(U\right)=-\frac{1}{kT}\left(\frac{\left\langle r_{ij}^{2}\exp\left[Fr_{ij}/kT\right]\right\rangle _{0}^{U}}{\left\langle \exp\left[Fr_{ij}/kT\right]\right\rangle _{0}^{U}}\right)+\frac{1}{kT}\left(\frac{\left\langle r_{ij}\exp\left[Fr_{ij}/kT\right]\right\rangle _{0}^{U}}{\left\langle \exp\left[Fr_{ij}/kT\right]\right\rangle _{0}^{U}}\right)^{2}
\]

\begin{equation}
\left[\frac{\partial^{2}}{\partial F^{2}}G_{F}\left(U\right)\right]_{F=0}=-\frac{1}{kT}\left\langle \Delta r_{ij}^{2}\right\rangle _{0}^{U}\label{Eq:FreeEnergyUF2}
\end{equation}
both evaluated at $F=0$. Combining these two, we have an approximation
for the free energy during force spectroscopy:

\begin{equation}
G_{F}\left(U\right)\approx G_{0}\left(U\right)-\left\langle r_{ij}\right\rangle _{0}^{U}F-\frac{1}{2}\frac{1}{kT}\left\langle \Delta r_{ij}^{2}\right\rangle _{0}^{U}F^{2}+\ldots+\mathrm{cnst.}\label{Eq:FreeEnergyUF_}
\end{equation}
These perturbation terms for the free energy in the force-free case
(i.e., $G_{0}\left(U\right)$) have a very simple meaning: During
force spectroscopy in the vicinity of $F=0$, the mean $\left\langle r_{ij}\right\rangle _{0}^{U}$
corresponds with the slope of $G_{F}\left(U\right)$, and the variance
$\left\langle \Delta r_{ij}^{2}\right\rangle _{0}^{U}$ corresponds
with the curvature of $G_{F}\left(U\right)$.

\vspace{2ex}
 
\begin{figure}[H]
\begin{centering}
\includegraphics[scale=0.5]{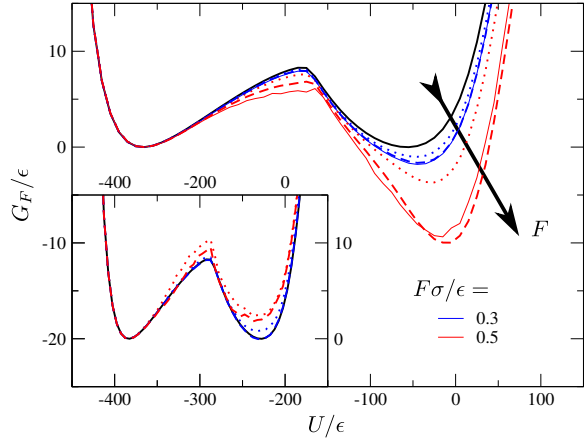} 
\par\end{centering}
\caption{\textit{\label{Fig:FreeEnergyUF_} }Approximations for the free energy
as a function of the ``contacts-number''. Besides the free energy
for $F=0$ (i.e., the black curve), there are two sets of colored
curves here corresponding with two different cases of force spectroscopy:
The blue lines are always associated with a force of $0.3\epsilon/\sigma$
on beads $\left\{ 43,43\right\} $ ($\Delta n=43$), and the red lines
are are always associated with a force of $0.5\epsilon/\sigma$ on
beads $\left\{ 15,15\right\} $ ($\Delta n=99$); note that the arrow
denotes an increasing value of the force, as well as of the relative
separation in index space. In the main panel, while all solid lines
are the exact free energies given by Eq.\ \ref{Eq:FreeEnergyUF},
the other curves represent their respective approximations via the
Taylor series of Eq.\ \ref{Eq:FreeEnergyUF_}: The dotted lines just
use the linear term, while the dashed lines also use the quadratic
term. Conversely, the auxiliary panel focuses on estimating the free
energy at $F=0$ by invoking Eq.\ \ref{Eq:FreeEnergyU0_}: Again,
the dotted and dashed lines respectively correspond with linear and
quadratic approximations. }
\end{figure}

Eq.\ \ref{Eq:FreeEnergyUF_} is completely general, and we now examine
its efficacy in describing the free energy of our polymer. Importantly,
considering the black curves in Fig.\ \ref{Fig:DistnaceMomentUF},
together with the black curve for $G_{0}\left(U\right)$ (the one
which frequently appears in our graphs), we have already presented
all the necessary ingredients for this computation. In the main panel
of Fig.\ \ref{Fig:FreeEnergyUF_}, we consequently present two sets
of perturbations: The blue curves are for a small perturbation (i.e.,
$F=0.3\epsilon/\sigma$ on beads which are near to each other with
$\Delta n=43$), and the red curves are for a large perturbation (i.e.,
$F=0.5\epsilon/\sigma$ on beads which are far from each other with
$\Delta n=99$). While the solid curves are for the exact free energies
by Eq.\ \ref{Eq:FreeEnergyUF}, the first-order and second-order
approximations, based on the Taylor expansion of Eq.\ \ref{Eq:FreeEnergyUF_},
are respectively given as dotted and dashed curves. Naturally, for
a fairly small force, just the slope at $F=0$ is sufficient, yet
for a fairly large force, also the curvature at $F=0$ is necessary.
In any case, we can state that by employing the two terms given in
the Taylor series above (i.e., in Eq.\ \ref{Eq:FreeEnergyUF_}),
we can generally capture the functionality of the free energy of our
polymer in force spectroscopy. 
\end{onehalfspace}

Nevertheless, in experiments of force spectroscopy, the opposite situation
is usually the case: Measurements are made at a finite $F$, and estimates
for $G_{0}\left(U\right)$ are desired. As such, two other Taylor
approximations are required for that, specifically, for the moments
of the pairwise distance which appear in Eq.\ \ref{Eq:FreeEnergyUF_}.
Following an analogous approach as in Eqs.\ \ref{Eq:FreeEnergyUF1}
and \ref{Eq:FreeEnergyUF2}, $\left\langle r_{ij}\right\rangle _{F}^{U}\approx\left\langle r_{ij}\right\rangle _{0}^{U}+\frac{1}{kT}\left\langle \Delta r_{ij}^{2}\right\rangle _{0}^{U}F$
and $\left\langle \Delta r_{ij}^{2}\right\rangle _{F}^{U}\approx\left\langle \Delta r_{ij}^{2}\right\rangle _{0}^{U}$.
Making the appropriate substitutions in Eq.\ \ref{Eq:FreeEnergyUF_},
together with some rearrangement, we obtain the following:

\begin{onehalfspace}
\begin{equation}
G_{0}\left(U\right)\approx G_{F}\left(U\right)+\left\langle r_{ij}\right\rangle _{F}^{U}F-\frac{1}{2}\frac{1}{kT}\left\langle \Delta r_{ij}^{2}\right\rangle _{F}^{U}F^{2}+\ldots+\mathrm{cnst.}\label{Eq:FreeEnergyU0_}
\end{equation}
Essentially, this expression is almost identical with Eq.\ \ref{Eq:FreeEnergyUF_},
except that one of the terms has its sign switched. 

We consequently present in the auxiliary panel of Fig.\ \ref{Fig:FreeEnergyUF_}
estimates for the free energy of the force-free scenario (i.e., the
black curve). The coloring here is analogous with that of the main
panel: The blue curves use the Zwanzig data of $F=0.3\epsilon/\sigma$
with $\Delta n=43$, and the red curves use the Zwanzig data of $F=0.5\epsilon/\sigma$
with $\Delta n=99$; again, the dotted and dashes lines respectively
correspond with the linear and quadratic approximations of Eq.\ \ref{Eq:FreeEnergyU0_}.
The approximations here are not quite as successful as those of the
main panel: Good agreement is only seen for small perturbations, and
even then we have to include both the linear as well as the quadratic
term in the expansion. This is perhaps due to the fact that extra
approximations were introduced for Eq.\ \ref{Eq:FreeEnergyU0_} (i.e.,
for $\left\langle r_{ij}\right\rangle _{F}^{U}$ and $\left\langle \Delta r_{ij}^{2}\right\rangle _{F}^{U}$).
Overall, this auxiliary panel of Fig.\ \ref{Fig:FreeEnergyUF_} conveys
the fact that if one is interested in estimating the force-free $G_{0}$
experimentally, one must generally aim for measurements at small forces,
as well as with small $\Delta n$, and we also strongly recommend
employing both terms in the Taylor series for a better approximation.

\vspace{2ex}

\end{onehalfspace}

\section{\textit{Activation Energy of the Crystal-Coil Transition \label{Sec:Activation}}}

We now finally connect our current equilibrium study for free energies
of generic polymers with the various kinetic studies of protein folding.
As alluded to earlier, we do so with one crucial assumption: The ``contacts-number''
is a decent approximation for the ideal reaction coordinate of the
unfolding process. We can presume this with relative confidence since
Refs.\ \citep{LeitoldDellago_JCP2014,LeitoldDellago_JPCM2015} found
that the ``contacts-number'' gives the best correspondence with
the committor probability (in comparison with many other order parameters),
and besides, many kinetic studies of protein folding do actually employ
the ``contacts-number'' for their reaction coordinate. 

\begin{figure}[H]
\begin{centering}
\includegraphics[scale=0.5]{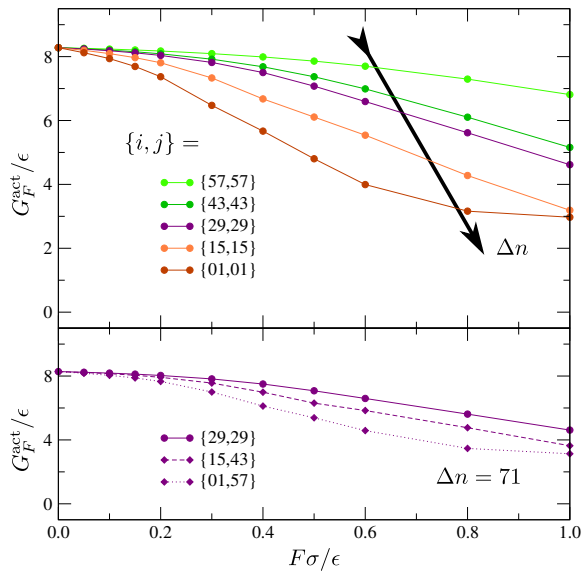} 
\par\end{centering}
\caption{\textit{\label{Fig:Activation} }The activation energy as a function
of the applied force. The coding here is reminiscent of Fig.\ \ref{Fig:FreeEnergyUij}.
In the top panel, the different colors correspond with various separations
between the bead pair; the arrow means increasing $\Delta n$; conversely,
the bottom panel maintains a constant $\Delta n$. The circles, connected
by solid lines, are for symmetric scenarios, and the diamonds, connected
by broken lines, are for asymmetric cases. }
\end{figure}

\begin{onehalfspace}
With this presumed choice for the reaction coordinate, the activation
energy $G_{F}^{\mathrm{act}}$ for the crystal-coil transition is
defined as follows:

\begin{equation}
G_{F}^{\mathrm{act}}=G_{F}\left(U^{\ddagger}\right)-G_{F}\left(U^{\mathcal{A}}\right)\label{Eq:ActivationF}
\end{equation}
We depict it schematically in Fig.\ \ref{Fig:FreeEnergyUF}; realize
that the locations of the extrema (e.g., $U^{\ddagger}$) are implicitly
dependent on $F$. By enumerating these points in our curves (e.g.,
those in Fig.\ \ref{Fig:FreeEnergyUF}), we evaluate the activation
energy, and we present it as a function of the applied force, in Fig.\ \ref{Fig:Activation}.
Each curve here is for a different bead pair in force spectroscopy,
with the top and bottom panels respectively being for symmetric and
asymmetric pulling. In terms of the dependence on the bead pair, we
make analogous observations as we made in the context of Fig.\ \ref{Fig:FreeEnergyUij}:
In the top panel, we note that the activation barrier decreases as
the relative separation (in index space) increases, and in the bottom
panel, we observe that the equivalent effect can be achieved by pulling
on a segment at the edge of the polymer rather than at the middle
of the polymer. Regardless, the striking signature of Fig.\ \ref{Fig:Activation}
is that all the different curves exhibit almost the same functionality
with $F$: Originating at the value of the activation energy for the
force-free scenario (i.e., $G_{0}^{\mathrm{act}}=8.28\epsilon$),
they all posses a negative slope and a negative curvature (at least
for moderate forces below $1\epsilon/\sigma$). Note that this is
reminiscent with our observations for the free energy over the entire
order parameter in the specific case of $\left\{ i,j\right\} =\left\{ 29,29\right\} $
in Fig.\ \ref{Fig:FreeEnergyUF}. \vspace{2ex}
 
\end{onehalfspace}

\begin{figure}[H]
\begin{centering}
\includegraphics[scale=0.5]{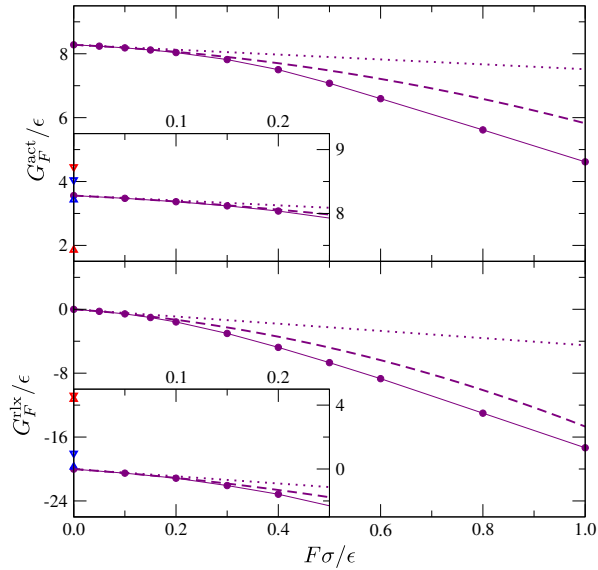} 
\par\end{centering}
\caption{\textit{\label{Fig:Activation_} }Estimates for changes in the free
energy as a function of the applied force. The top panel is for the
activation energy, and the bottom panel is for the relaxation energy.
Specifically for the bead pair $\left\{ 29,29\right\} $, the violet
circles are the exact data, with the line between them just serving
as a guide; the dotted and dashed lines correspond with the first-order
and second-order approximations, respectively, given by Eq.\ \ref{Eq:ActivationF_}.
The auxiliary panels focus on the curves in the vicinity of the force-free
scenario. Here, we also plot points on the ordinate-axis, which correspond
with estimates for the free energies at $F=0$: Specifically, the
down and up triangles respectively represent the linear and quadratic
approximations of Eq.\ \ref{Eq:Activation0_}; corresponding with
the Zwanzig curves of the auxiliary panel of Fig.\ \ref{Fig:FreeEnergyUF_},
the blue symbols are based on $F=0.3\epsilon/\sigma$ on $\left\{ i,j\right\} =\left\{ 43,43\right\} $,
and the red symbols are based on $F=0.5\epsilon/\sigma$ on $\left\{ i,j\right\} =\left\{ 15,15\right\} $. }
\end{figure}

Why do we observe this generic functionality in force spectroscopy
regardless of which monomeric pair $F$ is applied on? To answer this
question, we substitute the Taylor approximation of Eq.\ \ref{Eq:FreeEnergyUF_}
in the definition of Eq.\ \ref{Eq:ActivationF}, obtaining the following
series for the activation energy in terms of the applied force:

\begin{onehalfspace}
\begin{equation}
G_{F}^{\mathrm{act}}\approx G_{0}^{\mathrm{act}}-\left[\left\langle r_{ij}\right\rangle _{0}^{\ddagger}-\left\langle r_{ij}\right\rangle _{0}^{\mathcal{A}}\right]F-\frac{1}{2}\frac{1}{kT}\left[\left\langle \Delta r_{ij}^{2}\right\rangle _{0}^{\ddagger}-\left\langle \Delta r_{ij}^{2}\right\rangle _{0}^{\mathcal{A}}\right]F^{2}+\ldots\label{Eq:ActivationF_}
\end{equation}
Importantly, we find that the slope and the curvature of the activation
energy are given as the differences between the means and the variances,
respectively, evaluated at the appropriate extrema of $F=0$. Note
that this expression is in effect a variation on the recent extension
of the Bell model: The linear term is the main aspect of the original
expression, while the quadratic term is an improvement that accounts
for the ``compliance'' \citep{HuangBoulatov_PAC2010}. Specifically
for the bead pair $\left\{ 29,29\right\} $, we examine the efficacy
of this expression in the top panel of Fig.\ \ref{Fig:Activation_}.
The violet circles are identical with their counterparts in Fig.\ \ref{Fig:Activation};
the dotted and dashed curves are respectively the first-order and
second-order approximations based on the Taylor series of Eq.\ \ref{Eq:ActivationF_}.
As can be clearly noticed, with a negative slope and a negative curvature
across all $F$, the quadratic approximation captures the general
functionality of the exact curves very well (the linear approximation
is inherently deficient in doing so). Note that we observe analogous
trends for other bead pairs as well. 

Considering the approximation of Eq.\ \ref{Eq:ActivationF_}, the
functionality of the free energy for our biomimetic chain now makes
perfect sense. According to Fig.\ \ref{Fig:DistnaceMomentUF}, $\left\langle r_{ij}\right\rangle _{0}$
and $\left\langle \Delta r_{ij}^{2}\right\rangle _{0}$ are fairly
negligible in the folded state, yet they become relatively noticeable
across the transition region. This makes the corresponding differences
between the moments in Eq.\ \ref{Eq:ActivationF_} positive, thus
yielding a negative slope and a negative curvature for the curves
of Fig.\ \ref{Fig:Activation_}. We suspect that most proteins have
such a signature. This is because in general, the folded state is
always compact with a stiff structure, and upon disentanglement, the
overall size increases together with the respective fluctuations;
in other words, a distance between an arbitrary residue pair in a
certain protein usually possesses the $F=0$ trend of Fig.\ \ref{Fig:DistnaceMomentUF}.
Nevertheless, some proteins may exhibit a different functionality
for $G_{F}^{\mathrm{act}}$, particularly because some of their residues
have specific preferences for each other. For example, if the fluctuations
in $r_{ij}$ significantly diminish across the transition region (perhaps
because the two relevant residues must adhere to certain portions
of the protein so that unfolding occurs), force spectroscopy will
exhibit the functionality of the blue curve of Fig.\ \ref{Fig:Sketch}.
Realize that other peculiar trends in protein unfolding have been
noted as well (e.g., the activation energy initially increases at
$F=0$) \citep{Makarov_JCP2016}. In any case, a given protein may
only have a few special pairs that play a pivotal role in the crystal-coil
transition. Thus in most cases, if one applies a force on a random
pair of residues in a certain protein, the functionality of the red
curve in Fig.\ \ref{Fig:Sketch} is expected in the vicinity of $F=0$.

\vspace{2ex}
 
\end{onehalfspace}

Finally, as mentioned earlier, experimentalists typically make measurements
at a finite force, attempting to obtain an estimate for the free energy
at $F=0$. For such a task in the context of the activation energy,
we consequently formulate an approximation that is similar to Eq.\ \ref{Eq:ActivationF_}.
By performing analogous mathematical manipulations as done in the
context of Eq.\ \ref{Eq:FreeEnergyU0_}, we obtain the following
useful expression for force spectroscopy:

\begin{onehalfspace}
\begin{equation}
G_{0}^{\mathrm{act}}\approx G_{F}^{\mathrm{act}}+\left[\left\langle r_{ij}\right\rangle _{F}^{\ddagger}-\left\langle r_{ij}\right\rangle _{F}^{\mathcal{A}}\right]F-\frac{1}{2}\frac{1}{kT}\left[\left\langle \Delta r_{ij}^{2}\right\rangle _{F}^{\ddagger}-\left\langle \Delta r_{ij}^{2}\right\rangle _{F}^{\mathcal{A}}\right]F^{2}+\ldots\label{Eq:Activation0_}
\end{equation}
In the top panel of Fig.\ \ref{Fig:Activation_}, by invoking two
sets of the Zwanzig data, at $F=0.3\epsilon/\sigma$ on beads $\left\{ 43,43\right\} $
and at $F=0.5\epsilon/\sigma$ on beads $\left\{ 15,15\right\} $
, we respectively plot these estimates for the force-free scenario
as blue and red (upwards and downwards) triangles respectively for
the (first-order and second-order) approximations of Eq.\ \ref{Eq:Activation0_}.
Just as in Fig.\ \ref{Fig:FreeEnergyUF_}, the data points here do
not demonstrate sufficient replication capabilities. We suspect that
a decent amount of the error stems from the fact that the locations
of the extrema shift with $F$. In any case, Eq.\ \ref{Eq:Activation0_}
still holds promise for experimentalists in estimating the force-free
$G_{0}^{\mathrm{act}}$, if they can make measurements for sufficiently
weak forces. 

\vspace{2ex}

\end{onehalfspace}

\section{\textit{Relaxation Energy of the Crystal-Coil Transition \label{Sec:Relaxation}}}

\begin{onehalfspace}
Besides $G_{F}^{\mathrm{act}}$, which has an intimate connection
with the kinetics of unfolding, we can also evaluate the overall change
in the free energy between the collapsed and expanded states; we call
this the relaxation energy $G_{F}^{\mathrm{rlx}}$: We depict it schematically
in Fig.\ \ref{Fig:FreeEnergyUF}, and it is defined in an analogous
manner as the activation energy of Eq.\ \ref{Eq:ActivationF} (i.e.,
$G_{F}^{\mathrm{rlx}}=G_{F}\left(U^{\mathcal{B}}\right)-G_{F}\left(U^{\mathcal{A}}\right)$).
In turn, the corresponding approximation of Eq.\ \ref{Eq:ActivationF_},
as well as in Eq.\ \ref{Eq:Activation0_}, holds for any polymer:
The sole difference is that the $\ddagger$ index must be replaced
with the index of the unfolded state. Specifically for our polymer,
we present its relaxation energy, together with its relevant approximations,
in the bottom panel of Fig.\ \ref{Fig:Activation_}, using here the
same notation as in the respective top panel; this data is again for
the $\left\{ 29,29\right\} $ scenario. While this relaxation energy
is a strictly different property than the activation energy, the interesting
aspect in this graph is that it follows the same trend: $G_{F}^{\mathrm{rlx}}$
has a negative slope and a negative curvature as a function of the
applied force. For the overall free energy as a function of the ``contacts-number''
(e.g., the curves of $G_{F}\left(U\right)$ in Fig.\ \ref{Fig:FreeEnergyUF}),
we can consequently state that such a functionality is more or less
expected across the entire order parameter in the proximity of $F=0$.

Let us now discuss the ramification of these results. Foremost, an
expression reminiscent of Eq.\ \ref{Eq:ActivationF_} for the relaxation
energy has been derived in other works as well \citep{Makarov_JCP2016},
but there is a special implication with the observations in this work:
For a generic polymer that exhibits the crystal-coil transition with
no specific preference between its contacts, its relaxation energy
in force spectroscopy is always expected to exhibit a negative slope
and a negative curvature in the vicinity of $F=0$ (i.e., the red
curve of Fig.\ \ref{Fig:Sketch}). Of course, we made a similar statement
for the activation energy, but we make a further claim for $G_{F}^{\mathrm{rlx}}$:
We actually expect such a trend for the unfolding of any protein.
The reason for this stems in the moments of the bead-bead distance:
Comparing a very rigid crystal and a very floppy coil, both relevant
differences, between their means (i.e., $\left[\left\langle r_{ij}\right\rangle _{0}^{\mathcal{B}}-\left\langle r_{ij}\right\rangle _{0}^{\mathcal{A}}\right]$)
and between their variances (i.e., $\left[\left\langle \Delta r_{ij}^{2}\right\rangle _{0}^{\mathcal{B}}-\left\langle \Delta r_{ij}^{2}\right\rangle _{0}^{\mathcal{A}}\right]$),
will be strictly positive in the force-free limit (in consideration
of Fig.\ \ref{Fig:DistnaceMomentUF}), and thus, the functionality
observed in Fig.\ \ref{Fig:Activation_} is expected regardless of
the chemical peculiarities of the protein. Thus for the relaxation
energy in the vicinity of $F=0$, we cannot imagine the other options
discussed in the previous section for $G_{F}^{\mathrm{act}}$ (i.e.,
the blue curve of Fig.\ \ref{Fig:Sketch}); in other words, we expect
only one functionality type for it.

\vspace{2ex}

\end{onehalfspace}

\section{\textit{Conclusion \label{Sec:Conclusion}}}

In this work, with molecular simulations of the biomimetic polymer
of Taylor et al.\ \citep{TaylorBinder_PRE2009,TaylorBinder_JCP2009},
we have elucidated on the (generic) kinetic behavior of proteins in
force spectroscopy. In particular, we showed that the free energy
for the unfolding of an arbitrary chain (with no specific preferences
among its beads) exhibits a characteristic signature in the vicinity
of the force-free scenario: Above all, the activation energy has a
negative slope and a negative curvature as a function of the applied
force, and this is irrespective of the specific pair of monomers which
are being pulled (note Figs.\ \ref{Fig:Activation} and \ref{Fig:Activation_}).
While our molecular simulations are specific for the coexistence temperature
between the crystal and coil phases, our Zwanzig-based formalism is
completely general, and thus, we expect an analogous trend for any
temperature that exhibits an activation barrier between these two
metastable states. Regardless, the explanation for this behavior lies
in the Taylor approximation of Eq.\ \ref{Eq:ActivationF_}: The linear
and quadratic terms in the series are respectively given by the means
and the variances of the bead-bead distance, specifically evaluated
as a difference of the transition region with respect to the folded
state; in general, this distance is expected to have lower values,
with lower fluctuations, in the folded state, and this in turn yields
a negative slope and a negative curvature for the force functionality.
We also derived here Eq.\ \ref{Eq:Activation0_}, which is applicable
for any protein, and thus, it can be very useful for experimentalists:
By measuring the first and second moments of the bead-bead distance
at a finite force, one can make an approximation for the rate coefficient
of the (intrinsic) force-free folding. 

Considering that in reality residues have specific preferences for
each other, force spectroscopy may of course exhibit other trends,
especially if the pair being pulled has a unique role in the transition
mechanism (note Figs.\ \ref{Fig:Sketch}). Nevertheless, since the
behavior of heterogeneous biological polymers is significantly influenced
by the behavior of homogeneous generic polymers, we expect that the
observations in our work hold for most residues in most proteins.
This statement is in fact complementary with recent kinetic theories
which also expect such a behavior for most proteins \citep{Makarov_JCP2016}.
Note that this is in contrast with the Dudko formula, which presumes
that the bead-bead distance is always the reaction coordinate for
unfolding \citep{DudkoSzabo_PRL2006}. It has been suggested that
while such an assumption is formally valid in the vicinity of the
critical force at which the activation energy vanishes, it is usually
incorrect in the vicinity of the infinitesimal force; in essence,
there is a crossover in the reaction coordinate as the force increases
\citep{SuzukiDudko_PRL2010,ZhuravlevThirumalai_PNAS2016}. Specifically
in our work, we argue that the distance between the monomeric pair
cannot generally be the reaction coordinate in the force-free limit,
since such an order parameter cannot even cleanly distinguish between
the crystal and coil phases (note Fig.\ \ref{Fig:ProbabilityDistributionR}).
For the intrinsic folding process, the number of contacts a chain
makes with itself is a much better representation of the reaction
coordinate \citep{SunOnuchic_BJ2014,BerkovichBerne_JPCB2017}. This
is in fact the main assumption of our study, which is more or less
validated by the likelihood investigation of Refs.\ \citep{LeitoldDellago_JCP2014,LeitoldDellago_JPCM2015}.
On a final note, the mere fact that we attain reminiscent trends as
observed in kinetic studies of protein folding indicates that much
of force spectroscopy may be simply understood in terms of equilibrium
phenomena of generic polymers. Nevertheless, keep in mind that there
are some aspects of protein folding which cannot be described by our
generic chain: For example, intermediate states in the transition
mechanism are typically observed during force spectroscopy for heterogeneous
polymers, yet this aspect is absent for our homogeneous polymer \citep{HyeonThirumalai_PNAS2009,BerkovichFernandez_BJ2010}.
Besides, in describing the entire phase transition correctly for purposes
of force spectroscopy, one must be also mindful of the kinetics associated
with the experimental apparatus itself  \citep{HinczewskiNetz_PNAS2010}. 

\begin{onehalfspace}
\vspace{2ex}

\end{onehalfspace}

\pagebreak{} 

\section*{\textit{\normalsize{}Conflicts of Interest \label{Sec:Conflict}}}

There are no conflicts to declare.

\begin{onehalfspace}
\vspace{4ex}

\end{onehalfspace}

\section*{\textit{\normalsize{}Acknowledgments \label{Sec:Acknowledgments}}}

We are grateful to the Alexander von Humboldt Foundation and the National
Science Foundation for funding most of this work. We also appreciate the
computer cluster of Baron Peters, as well as that of Mark Santer. Besides,
we acknowledge several discussions with Kurt Kremer.

\begin{onehalfspace}
\vspace{4ex}

\end{onehalfspace}

\section*{\textit{\normalsize{}Bibliography \label{Sec:Bibliography}}}

\begin{onehalfspace}
\renewcommand{\section}[2]{}

{\small{}\bibliographystyle{unsrt}
\bibliography{Manuscript}
 }{\small\par}

\vspace{2ex}
 
\end{onehalfspace}

\end{document}